\documentclass[twocolumn,showpacs,pra,preprintnumbers,amsmath,amssymb]{revtex4}

 \usepackage[pdftex]{graphicx}
\usepackage{dcolumn}
\usepackage{bm}
\usepackage{color}

\definecolor{DarkGreen}{rgb}{0,0.7,0.08} 
\definecolor{Grey}{rgb}{0.5,0.5,0.5}
\definecolor{Red}{rgb}{0.8,0.3,0.3}
\definecolor{Blue}{rgb}{0.2,0.2,1.0}

\newcommand{\newtext}[1]{#1}

\begin{document}


\title{Structure, Bose-Einstein condensation and superfluidity\\ of {two-dimensional} confined dipolar assemblies}
\author{Piyush Jain}
\email{jain@ualberta.ca}
\affiliation{Department of Physics, University of Alberta, Edmonton, Alberta, Canada}
\author{Fabio Cinti}
\email{cinti@ualberta.ca}
\affiliation{Department of Physics, University of Alberta, Edmonton, Alberta, Canada}
\author{Massimo Boninsegni}
\affiliation{Department of Physics, University of Alberta, Edmonton, Alberta, Canada}
\date{\today}                                           

\begin{abstract}
Low temperature properties of harmonically confined two-dimensional assemblies of dipolar bosons
are systematically investigated by Monte Carlo simulations. Calculations carried out for different numbers of particles and strengths of the confining potential yield evidence of a quantum phase transition from a superfluid to a crystal-like phase, consistently with what is observed in the homogeneous system.  It is found that the crystal phase nucleates in the center of the trap, as the density increases. 
Bose-Einstein condensation  vanishes at $T = 0$ upon entering the crystalline phase, concurrently with the disappearance of the superfluid response.
   
\end{abstract}

\maketitle


\section{Introduction}\label{sect:intro}

Bose-Einstein Condensation (BEC) and superfluidity (SF) are intimately connected quantum many-body phenomena \cite{Leggett08}, both underlain by long cycles of exchange of identical particles \cite{Feynman1953}. In three dimensions BEC and SF appear simultaneously, and become almost synonymous when interactions among particles are weak, a condition normally experimentally  attained in trapped  ultracold gases \cite{Davis1995, Anderson1995}. In that limit, the fractions of the system that are superfluid and Bose {condensed} both approach 100\% at the temperature $T\to 0$. This provides a major justification for theoretical studies based on  mean-field  techniques \cite{Krauth1996,Heinrichs1998}. 

On the other hand, the effect of strong inter-particle interactions on SF and BEC is very different. A chief example is provided by liquid $^4$He, which approaches the 100\% superfluid limit as $T\to 0$, while its condensate is largely depleted; indeed, theoretical \cite{moroni04,Boninsegni2006b} and experimental \cite{azuah,azuah2} estimates for the condensate fraction  in superfluid $^4$He yield a value in the neighbourhood of 7-8\%, at  $T$=0. Furthermore, no theoretical understanding currently exists of the relationship, if any, between the local values of the superfluid and condensate fractions. Broadly speaking, the most quantitative aspects of the relationship between these two quantities remain to be further elucidated.

Impressive advances in the trapping and manipulation of ultracold atoms, appear to offer a pathway to a more in-depth investigation of this relationship, by making it feasible to create in the laboratory ``artificial" many-body systems characterized by interactions of  variable strength and shape. Thus, one may smoothly interpolate between weakly and strongly interacting regimes, as well as explore the effect on SF and BEC of interactions not (easily) realized in ordinary condensed matter systems. 

In recent times, attention has turned to the case of dipolar bosons, further bolstered by the realization of  Bose-Einstein condensation of Chromium \cite{Griesmaier2005}. The long-range and anisotropic nature of the dipole-dipole interaction leads to many fascinating phenomena \cite{Lahaye2009}. The conceptually simplest scenario is that of an assembly of dipolar bosons confined to two dimensions, with their dipole moments aligned by an external (electric or magnetic) field {directed perpendicular to the confinement plane}. In this case, the interaction between any two particles is purely repulsive, proportional to the inverse cubic power of the distance between them. 

Recently, it was shown that a mixture of equal-mass dipolar isotopes, in such a configuration, demixes at finite temperature due to quantum statistical effects \cite{Jain2011}.
{A  promising direction for implementing trapped dipolar assemblies, is through the use of Rydberg-excited atoms. The large induced dipole moments of such systems may facilitate the realization of solid-like phases \cite{Anderson1998,Pupillo2010}.}
\\ \indent
The phase diagram of {a homogeneous} system of purely repulsive dipolar bosons in two dimensions, has been explored by Monte Carlo simulations \cite{Buchler2007, Astrakharchik2007, Mora2007}. These works have yielded evidence of a quantum phase transition at $T$=0 between a superfluid and a crystal phase, but the quantitative characterization of such a phase transition, e.g., its location and the width of the coexistence region, has proven surprisingly difficult, due to the long-range nature of the interaction. Theoretical arguments \cite{spivak} have been put forth to the effect that a  conventional first-order quantum phase transition ought not occur in this system, as the long-ranged nature of the interactions renders the coexistence of two phases of difference densities, separated by a single macroscopic interface, energetically unfavourable. Rather, an ``emulsion" should form, consisting of  (relatively) large solid domains embedded in the superfluid. Such a scenario has not yet been observed in computer simulations, presumably due to the need of studying systems of size not currently attainable.
\\ \indent
The superfluid transition at finite temperature has been predicted to be compatible with the Berezinskii-Kosterlitz-Thouless (BKT) theory \cite{Filinov2010}. And, while no evidence has been seen of a \newtext{commensurate} {\it supersolid} phase \cite{spivak, Mora2007, Kurbakov2010}, its presence has not been ruled out yet (it is in fact predicted to occur in a system of dipolar bosons in the so-called Rydberg-blockaded regime \cite{cinti}). 
\\ \indent
It is expected that many of the outstanding issues will be soon elucidated by experiments on spatially confined assemblies of dipolar particles. Two-dimensional confinement can be achieved in the laboratory by means of a  harmonic trap in the transverse direction, of frequency $\omega_z$. Clearly, the system  must also be confined in the remaining two dimensions, as the interactions are purely repulsive, and this is accomplished by means of a second, in-plane harmonic trap, of frequency $\omega << \omega_z$,  realizing the so-called {\it pancake} geometry. The number of particles that can be currently trapped is typically of the order of a few thousands, which means that the physics of such a system is strongly influenced by its finite size. Thus, theoretical studies of  {\it finite} two-dimensional dipolar systems in a harmonic potential seem especially opportune and timely, in order to guide in the interpretation of experiments aimed at inferring bulk properties.
\\ \indent
However, a finite system of dipolar bosons is of interest in its own right, chiefly because novel fundamental understanding of the relationship between BEC and SF can be gained. 
\newtext{Indeed, theoretical work on clusters of $^4$He \cite{lewart,Sindzingre1989} has shown that superfluid and condensate fractions can be meaningfully and rigorously defined for systems comprising as few as $N$=70 particles. One should note that, for the two-dimensional systems considered in this work, because the system is finite, reduced dimensionality does not imply absence of BEC at finite temperature \cite{Bagnato1991,Hadzibabic2008}.}
Moreover, while superfluid and condensate are {\it uniform} in the bulk system, owing to translational invariance, in confinement they can display non-trivial {\it local} variations. Of particular import seems the comparison of the local superfluid and condensate fractions, which can lead to original insight into their interplay.
\\ \indent
{Previously, Lozovik {\it et al.} \cite{Lozovik2004}, and more recently Pupillo {\it et al.} \cite{Pupillo2010}, have studied harmonically confined dipolar systems of relatively small size (up to $N=40$ particles).
 In this paper, we investigate by means of Monte Carlo simulations the low temperature phase diagram of dipolar Bose assemblies comprising up to $N=1000$ particles, a number which appears relevant to current (or planned) experiments. \\ \indent 
The purpose of this work is twofold. On the one hand, we compare the ground state phase diagram of the trapped system against the predictions made in Refs. \cite{Buchler2007,Astrakharchik2007, Mora2007} for the bulk case. Our aim is to make contact between the theoretical phase diagram of the bulk and the quantitative information that can be provided by experiments on finite-size systems, comprising a relatively small number of particles.
Secondly, we compute both global and local densities of superfluid and Bose{-Einstein} condensate, based on numerically unbiased estimators, and study how the two quantities evolve as the strength of the interactions in the system increases (as measured by the density in the middle of the trap). 
To the best of our knowledge, this is the first system for which \newtext{both the local superfuid and condensate densities} are computed simultaneously for a strongly interacting many-body system, based on a methodology free from approximations.
\\ \indent
This paper is organized as follows. In section~\ref{sec:model} we introduce the model Hamiltonian for a two-dimensional system of dipolar Bose particles harmonically confined, and  discuss the relevant observables pertinent to structure, Bose-Einstein condensation and superfluidity. {We illustrate our results in the following three sections, Section~\ref{sec:phasediagram} focusing on the structure of the system with respect to system size and in-trap mean inter-particle distance, Section~\ref{sec:sfbec} focusing on the superfluidity and condensation, notably both global and local properties, and finally Section~\ref{sec:phasediagram2} discusses the ground state phase diagram, as well as the momentum distribution. We outline our conclusions in Section~\ref{sec:conclusions}.}



\section{Model and methodology}\label{sec:model}

We consider a two-dimensional system comprising  $N$ bosons of mass $m$, confined  by a harmonic potential of frequency $\omega$. Particles then interact via a purely repulsive, pair-wise dipolar potential, given by  $v(r) = D/r^3$, $r$ being the distance between the particles. The simplest way of realizing such an interaction, is with particles possessing an electric {(magnetic)} dipole moment, aligned along the transverse direction by an applied electric {(magnetic)} field \cite{Buchler2007}. The quantum-mechanical many-body Hamiltonian is given in dimensionless form by:
\begin{eqnarray}\label{hamiltonian}
\hat{H} =   -\frac{1}{2} \sum_i^N \nabla_i^2 + \Gamma \sum_i^N {\bf r}_i^2 + \sum_{i < j}^N \frac{1}{|{\bf r}_i - {\bf r}_j|^3}
\end{eqnarray}
where  ${\bf r}_i$ is the position of the $i$th particle, and where we have introduced length and energy scales respectively as $r_\circ = mD / \hbar^2$ and $\epsilon_\circ = \hbar^2 / m r_\circ^2$.
The trap strength parameter is $\Gamma = (1/2) (r_\circ / L)^4$, $L= \sqrt{\hbar / m \omega}$  being the oscillator length. It is also useful to introduce a \emph{reduced temperature} $T^* = {k_B} T/ (\hbar \omega)$, with $T$ expressed in units of $\epsilon_\circ$.

\begin{figure}[!tp]
\centering
\includegraphics[scale=0.25]{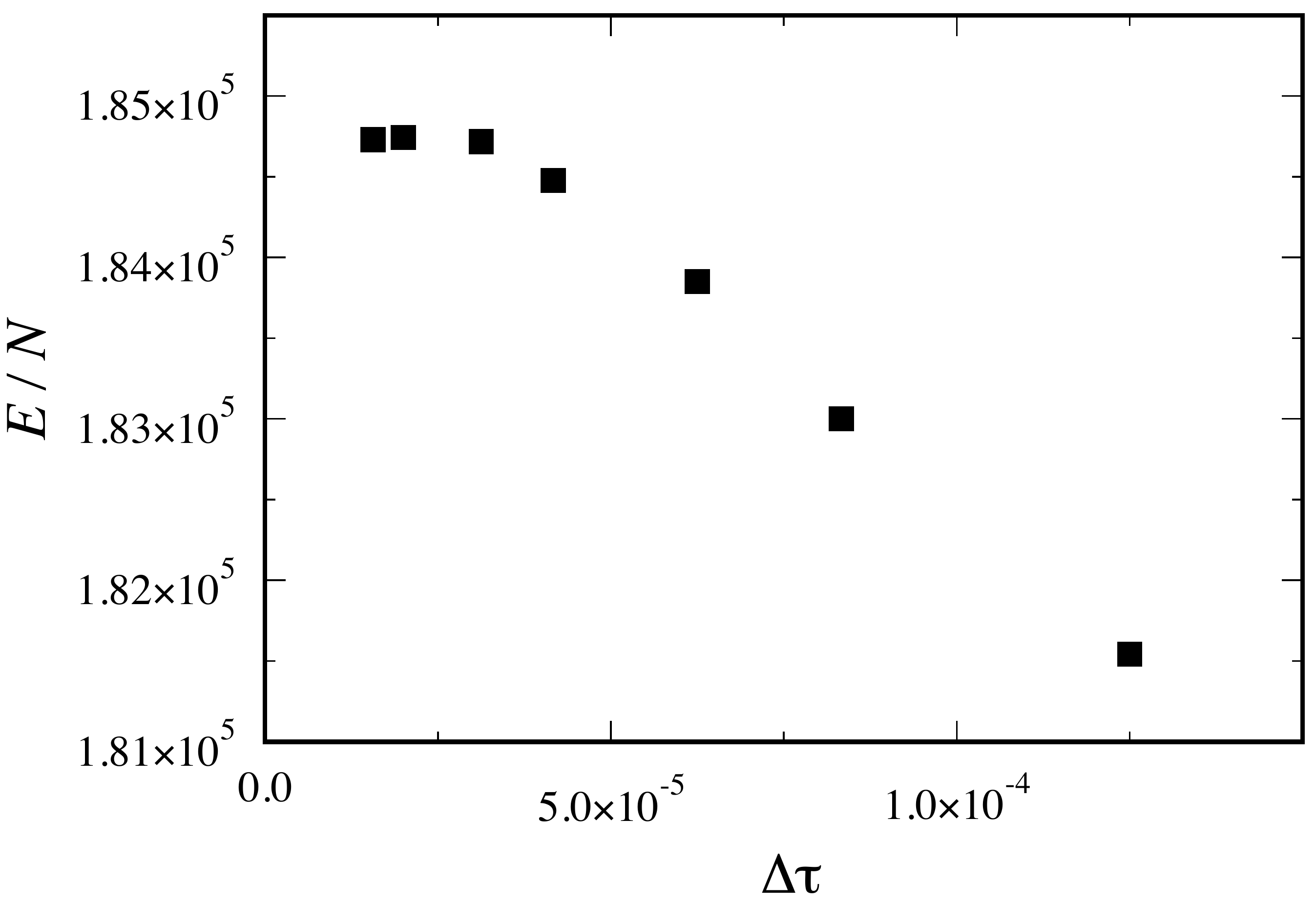}
\caption{Energy per particle as a function of (imaginary time) step size for Worm Algorithm simulations with $\Gamma = 5\times10^6$, $N=$ and inverse temperature $\beta = 1/500$ (where $k_B = 1$). In this case numerical convergence is found for $\Delta \tau = \beta/M \lesssim 2 \times 10^{-5}$ where $M$ is the number of  time steps.  Errors are smaller than the symbol size. }
\label{energytau}
\end{figure}


We investigate the low temperature properties of this finite system by means of Monte Carlo simulations.  We consider systems with particle numbers $N$ ranging from 20 to 1000, and trap strengths $\Gamma = 5 \times 10^{-2} - 5\times 10^6$. We use the Worm Algorithm, in the continuous-space path integral representation  \cite{Boninsegni2006,Boninsegni2006b}. The methodology is numerically exact, errors being only statistical in character, and reducible to an essentially negligible size with a relatively modest computer time expenditure.  Technical details of the calculation are the same as in other studies \cite{Mezzacapo2006,Mezzacapo2007,cuervo}; the use of the dipolar interaction in (\ref{hamiltonian}) entails no particular difficulty. 
All of our quoted ground state results, with their uncertainties, are obtained by extrapolating our numerical estimates to the limit of temperature $T=0$. In general, we have observed that the structural and energetic properties of the assemblies considered here do not change, within the statistical error of our calculations, if the temperature $T^\star$ is $\lesssim$ 0.1, below which the system may be regarded as essentially in its ground state.  As customary in this type of numerical study, we have also carried out extrapolation of our estimates to the limit of vanishing time step \cite{cuervo}. \newtext{As a typical example, Fig.~\ref{energytau} shows the total (potential plus kinetic) energy per particle as a function of (imaginary time) step size $\Delta \tau$. As the step size is made sufficiently small, the energy converges to a fixed value.}
\\ \indent
As mentioned above, one of our aims is to make contact between the ground state properties of the finite assembly considered here, and those of the same system in the bulk limit, where the physics is determined solely by the mean inter-particle distance $r_s = 1/\sqrt{\bar\rho\ r_\circ^2}$, $\bar\rho$ being the particle density.  For the trapped system, the density is not uniform, but rather is a function $\rho(r)$ of the radial distance from the centre of the trap. Near the centre of the trap, the physical behaviour of the finite system ought resemble most closely that of the bulk; thus, we associate a value of $r_s$ to the trapped system, by extrapolating to $r\to 0$ the behaviour of the function 
\begin{equation}\label{rs}
r_s(r)=\frac{1}{r_\circ \sqrt{\rho(r)}}
\end{equation}
Details of the extrapolation are given below.\\ \indent
Besides the radial density $\rho(r)$, we compute the local superfluid ($\rho_s(r)$), and condensate ($\rho_c(r)$) densities. The average superfluid density $\rho_s$ is computed using the area estimator \cite{Sindzingre1989}. The local superfluid density is computed by means of an estimator first proposed by Kwon \emph{et al.}  \cite{Kwon2006}:
\begin{eqnarray}
\frac{\rho_s(r)}{\rho(r)} = \frac{4 m^2 T \langle A A(r) \rangle}{\beta \hbar^2 I_c(r)}
\label{eq:lsd}
\end{eqnarray}
where $\beta=1/T$ (we set the Boltzmann constant equal to one), $\langle ...\rangle$ stands for thermal average, $A$ is the area swept by the many-particle in the $x-y$ plane, and where $A(r)$ and $I_c(r)$ are respectively the contributions to $A$ from paths in a shell of radius $r$, and $I_c(r)$ its classical moment of inertia. 

The estimation of the condensate fraction is carried out based on the ideas expounded in Refs. 
\cite{Lowdin1955,Onsager1956}. Specifically, we compute by Monte Carlo the angularly averaged one-body density matrix
\begin{equation}
\rho_0(r,r^\prime) = \langle \hat\psi^\dagger(r^\prime)\hat\psi(r)\rangle
\end{equation}
where $\hat \psi$ and $\hat\psi^\dagger$ are the usual Bose field operators. The matrix $\rho_0$ has positive-definite eigenvalues, the largest of which is the average number of particles in the maximally occupied natural orbital (condensate), which is the corresponding eigenfunction \cite{cazzate1}. The assumption is made here that the condensate should be cylindrically symmetric. As we are only interested in its largest eigenvalue $N_0$ here, and its associated natural orbital $\phi_0$, we solve the eigenvalue equation for $\rho_0$
iteratively (numerically), using the power method. The condensate fraction is then given by $N_{0}/N$ with corresponding local condensate density given by $\rho_c(r) = |\phi_{0}(r)|^2$.

\section{Structure}\label{sec:phasediagram}

\begin{figure}[!t]
\centering
\includegraphics[scale=0.35]{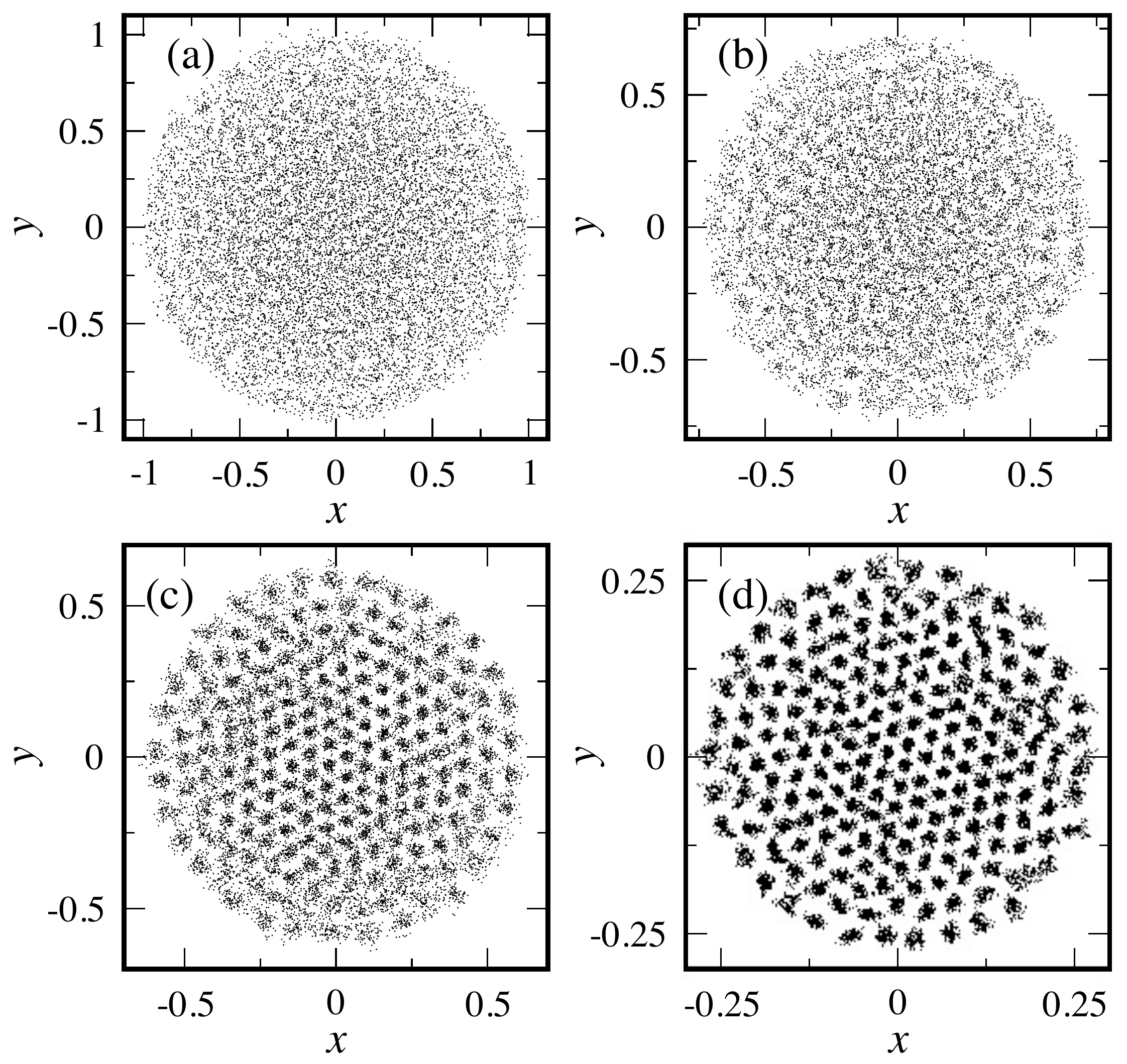}
\caption{Snapshots of many-particle world-lines for a trapped system comprising $N=200$ bosons, in the low temperature limit (i.e., $T\to 0$), for values of the trap strength
(a) $\Gamma = 1\times 10^4$\,;
(b) $\Gamma = 5\times 10^4$\,; 
(c) $\Gamma = 1\times 10^5$\,;
(d) $\Gamma = 5\times 10^6$\,.}
\label{figure1b}
\end{figure}

We begin by discussing the structural properties of the system as a function of the density of particles in the trap, in the $T\to 0$ (ground state) limit. Numerical studies of the ground state phase diagram of a two-dimensional homogeneous system of dipolar bosons have yielded evidence of a superfluid-to-crystal quantum phase transition \cite{Buchler2007, Astrakharchik2007, Mora2007}. The precise determination of the location and width of the coexistence region has so far proven rather challenging. B\"uchler {\it et al.} \cite{Buchler2007} place it within the relatively large range $0.045 \le r_s\le 0.065$, and other estimates \cite{Astrakharchik2007,Mora2007,Filinov2010} are in general agreement. The question is what remnants, if any, of such a phase transition can be observed on a finite system.
\\ \indent
A first, general comment is in order: one must exercise caution when referring to \emph{phase transitions} in finite-size (mesoscopic) systems. It is well known that, although such systems may exhibit ``phase-like behaviour'', the coexistence of two phases, which is typical of a first-order phase transition, is energetically disfavoured by the presence of an interface separating the two phases. Thus, in a finite-size system, especially of the size treatable in any realistic numerical simulation, phase transitions are smoothed out \cite{Fisher1982}.  To some extent, therefore, even though we shall make use of this terminology, any reference to ``phase" must be understood qualitatively. As well,  obviously the fact that the system is confined by a {\it harmonic} trap can be expected to influence qualitatively its physics, especially at the border where particle localization is enhanced by the steepness of the confining harmonic potential. 

\begin{figure}
\centering
\includegraphics[scale=0.4]{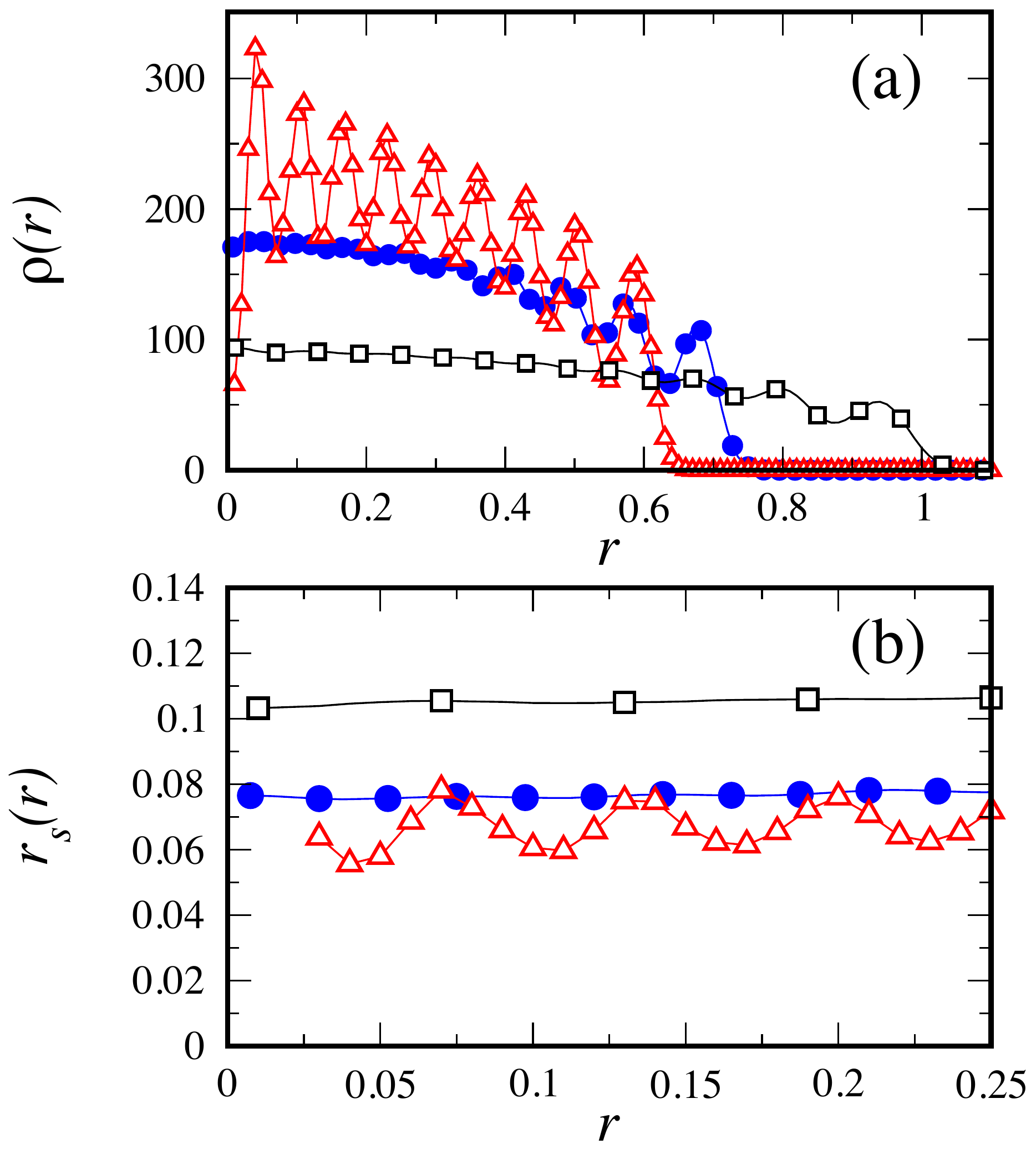}
\caption{(Color online) Density profiles for a dipolar system comprising $N$=200 particles, for values of the trap strength $\Gamma=1.0\times 10^4$ (squares),  $\Gamma=5.0\times 10^4$ (circles),
$\Gamma=1.0\times 10^5$ (triangles). All values of $\Gamma$ are in units of $\epsilon_\circ$. {\it Upper panel}: Local density $\rho(r)$. {\it Lower panel}: Corresponding local value of the inter-particle distance $r_s(r)$, defined in Eq. (\ref{rs}). Statistical errors are smaller than symbol sizes.
}
\label{figure1}
\end{figure}

Figure \ref {figure1b} shows instantaneous configurations (i.e., particle world lines) of a system comprising $N$=200 dipolar bosons, for different values of the trap strength $\Gamma$, increasing from (a) to (d). These results corresponds to temperatures sufficiently low that the system may be regarded essentially in its ground state. These snapshots offers visual, qualitative insight into the evolution of the system as the density inside the trap is increased. 
The presence of a featureless fluid phase at low density, and the emergence of crystalline order at high density, with particles arranged on a triangular lattice near the centre of the trap, are clear in Figure \ref {figure1b}. Specifically, at high density (i.e., greater trap strength) particle world lines become localized, and exchanges of (indistinguishable) particles suppressed, as shown for the largest trap strength in 
Figure ~\ref{figure1b}(d).

More quantitative information is offered by plots such as those of  Figure \ref{figure1}, which shows radial density profiles for a trapped system comprising $N$=200 particles, for different values of the confining strength $\Gamma$ of the trapping potential. The upper panel of the figure shows the local density of particles $\rho(r)$. At low density (i.e., low values of $\Gamma$), $\rho(r)$ is essentially constant near the centre, whereas at the edge of the trap, where the confining potential dominates the energetics, a shell structure forms. On the other hand, at high density  the dipole-dipole interaction energy dominates in the centre of the trap, where $\rho(r)$ displays oscillations that are consistent with a regular arrangement of particles.
\\ \indent
On the expectation that the physical behaviour of the system should mimic that of the bulk near the centre of the trap, we may compare results for a trapped system with those in the thermodynamic  
limit by attributing an effective value of $r_s$ to the confined system. As explained above, we do so by taking the value of $r_s(r)$ at $r=0$ (i.e., the centre of the trap), or, in case of oscillatory behaviour (e.g., triangles in Figure \ref{figure1}(b)), we take the value of $r_s(r)$ averaged over few periods of oscillation near the centre. The lower panel of Figure \ref{figure1} shows typical results for $r_s(r)$, as defined as in (\ref {rs}). We estimate the uncertainty in the values of $r_s$ determined in this way, quoted in the remainder of the paper, to be of the order of 5\%.
\\ \indent
For a system with $N$=200 particles, the results shown in Figure \ref{figure1}  suggest that a transition from a uniform liquid phase to a crystalline one, takes place for $r_s$ roughly below $r_s \lesssim 0.06$. A more precise assessment of the density region within which the transition takes place, can be obtained by studying superfluidity and Bose-Einstein condensation; we come back to this later on.
\begin{figure}[!t]
\begin{center}   
\includegraphics[scale=0.35]{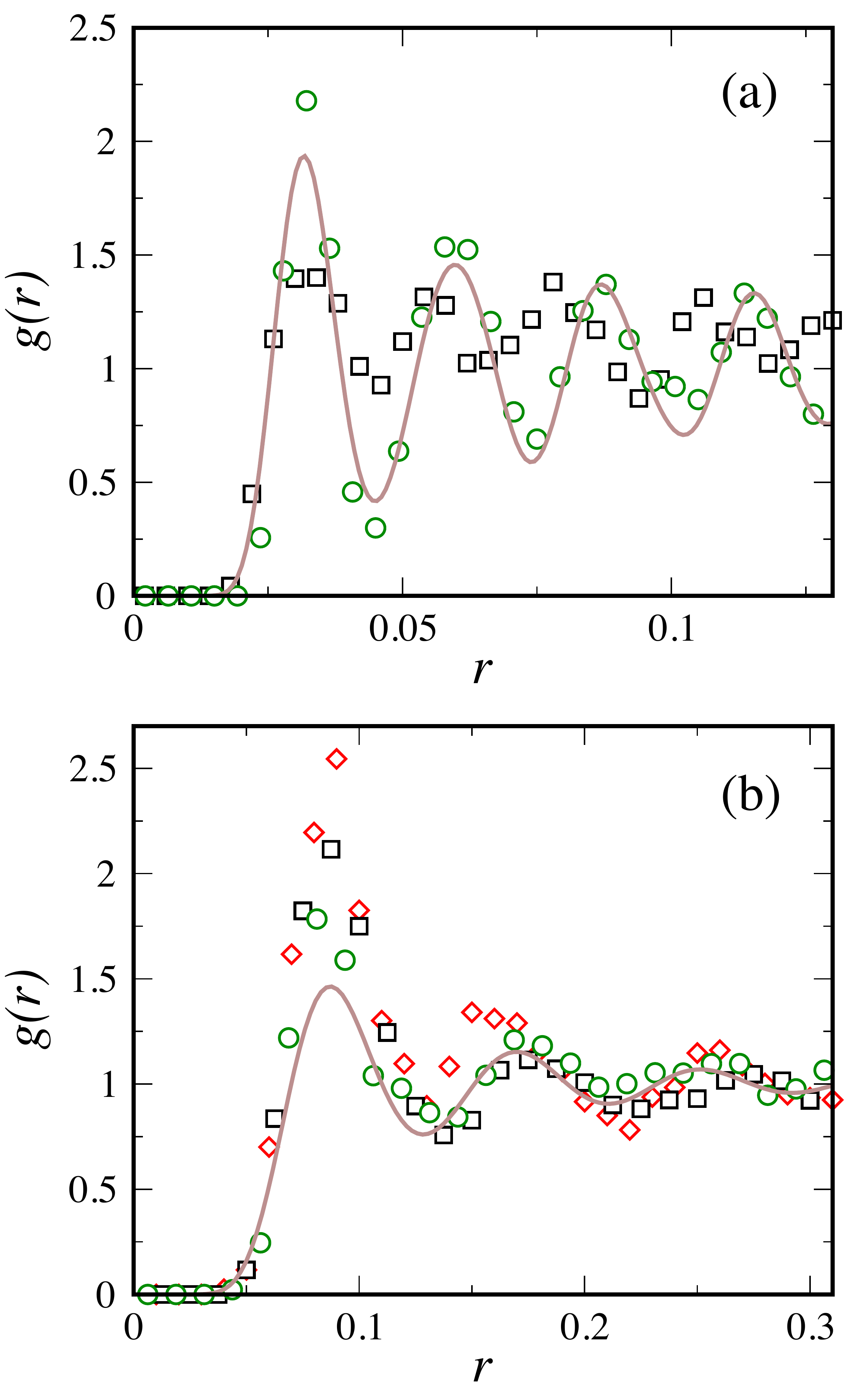}
\end{center}   
\caption{(Color online) Local pair correlation function for trapped dipolar bosons with $N$=100\,(diamonds), 200\,(squares), and 1000\,(circles), and in the bulk case (solid lines). The inter-particle distance $r_s$ in the middle of the trap is  $r_s= 0.03$ (a) and {$r_s = 0.085$} (b). Statistical errors are of the order of the symbol size.}
\label{localgr}
\end{figure}
\begin{figure}[!t]
\centering
\includegraphics[scale=0.5]{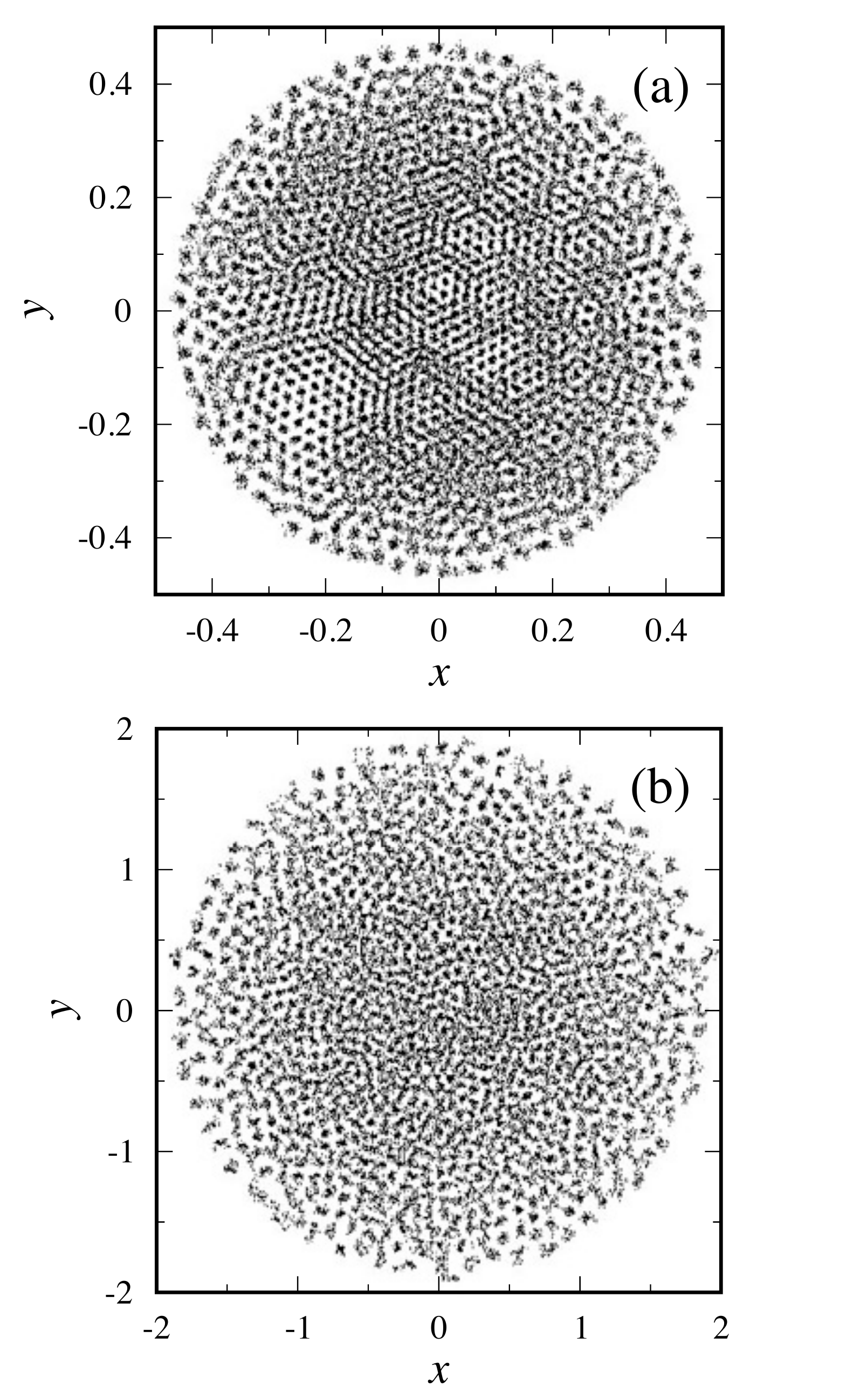}
\caption{Snapshots of many-particle world-lines for a trapped system comprising $N=1000$ bosons, in the low temperature limit (i.e., $T\to 0$), for values of the trap strength $\Gamma$ yielding an interparticle distance in the middle of the trap equal to
(a) $r_s = 0.03$\,;
(b) $r_s = 0.085$.}
\label{n1000config}
\end{figure}

\indent In order to establish a more quantitative connection between the physics of the trapped assembly and that 
of a bulk system, we compute the pair correlation function in
a circular region of radius $R$, centered at $r=0$. 
Such a ``local pair correlation function" inside this region, $g_R(r)$, has the normalization
\begin{eqnarray}
2\pi\rho_R \int^R_0 \ dr\ r\  g_R(r)  = N_R - 1
\end{eqnarray}
where $N_R$ is the average number of particles in the bounded region, and the 
corresponding average density is given by $\rho_R = N_R / \pi R^2$. \\ \indent
Figure \ref{localgr} shows the computed $g_R(r)$ for trapped systems with different number of particles, specifically  $N$=100\,(diamonds), 200\,(squares), and 1000\,(circles), in the $T\to 0$ limit. Results are shown for different trap strengths $\Gamma$, adjusted to yield a value of $r_s$ in the middle of the trap  around 0.03 (top panel) and 0.085 (bottom panel). Here, we took $R=0.13$ and 0.3 respectively, i.e., values smaller than the radius at which the density goes to zero in the trap.
For $r_s = 0.03$, the ground state of the bulk system is crystalline, a superfluid for $r_s = 0.085$.  Both figures also show (solid lines) the corresponding pair correlation functions for the bulk system with the same values of $r_s$, also computed in the $T$$\to$ 0 limit (a separate calculation was performed for this case, using standard bulk methodology, with periodic boundary conditions).
\\ \indent
The first observation is that in both cases, on increasing the number of particles $g_R(r)$ approaches that of the bulk system, as expected. Moreover, the finite size of the system, combined with the fact that the trap is harmonic, enhances particle localization and therefore strengthens crystalline order. Results shown in Figure \ref{localgr}(a) correspond to $r_s = 0.03$ in the middle of the trap; $g_R(r)$ displays marked oscillations for both $N$=200 and 1000 particles. However, for the smaller system size the height of the peaks is lower, and the distance between successive peaks is also different from what observed in the bulk. This suggests that, although the physics of the system is solid-like in both cases, for the smaller system it is substantially influenced by the trap, whereas with five times more particles it is already rather close to that of the bulk. Indeed, the pair correlation function in the center of the trap is virtually indistinguishable from that of the bulk system beyond the first peak, within our statistical errors. It displays the pronounced oscillations that one expects for a crystalline phase.
\\ \indent
Figure ~\ref{localgr}(b) shows results for the case  $r_s = 0.085$. Here too, the local pair correlation function has a noticeable dependence on the size of the system. However, in this case the distance between the peaks is roughly the same in all cases, i.e., the influence of the edge on the physics of the system at the center of the trap is less significant than in the case previously discussed (not surprisingly, as the trap is weaker). As the size of the system is increased from $N=100$ to $N=1000$ particles, the most significant change is the height of the first peak of $g_R(r)$, which decreases by roughly 30\%.  Beyond the first peak, the pair correlation functions for $N$=200 and $N$=1000 are essentially indistinguishable from one another, as well as from that of the bulk system (again within the statistical uncertainties of our calculation), and are consistent with liquid-like behavior.

\begin{figure}[t]
\begin{center}   
\includegraphics[scale=0.3]{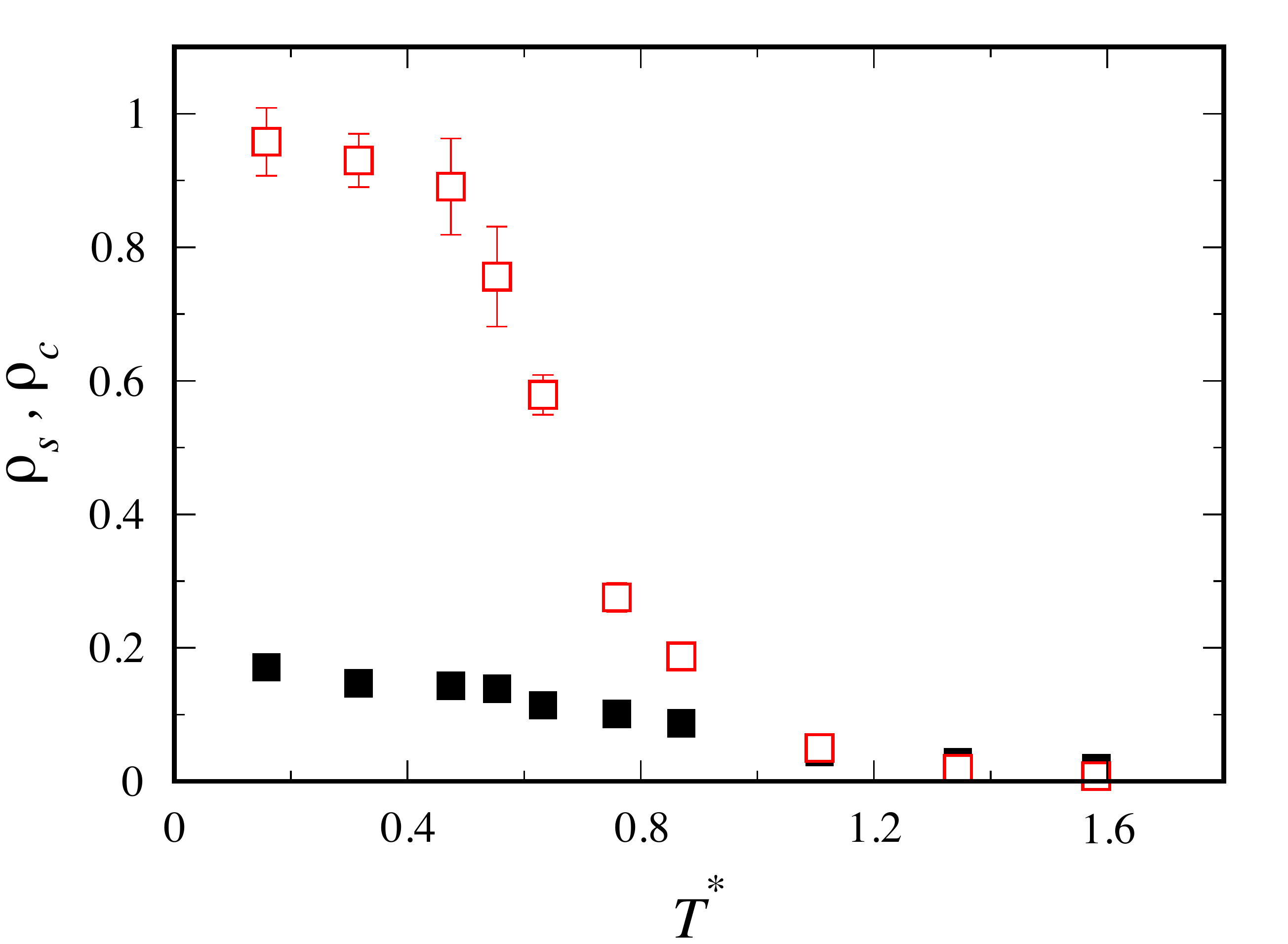}
\end{center}   
\caption{(Color online) Superfluid fraction $\rho_s$ {(red open squares)} and condensate fraction $\rho_c$ {(black filled squares)} as function of reduced temperature for system with $N=100$ and trap strength {yielding a value of the inter-particle distance $r_s = 0.22$ in the center of the trap}.}
\label{fig:rhosrhoc}
\end{figure}

\indent
{Figure \ref{n1000config} shows snapshots as in Figure \ref {figure1b} for $N=1000$ at low temperature, in correspondence of  two values of $r_s$ shown in Fig.~\ref{localgr}. For $r_s = 0.03$, an orderly arrangement of atoms in a triangular lattice occurs in the center of the trap, confirming the picture given by consideration of the local pair correlation function.}  It is important to note, however, that exchanges still take place, primarily in the peripheral disordered region, where the system goes from bulk crystal (center of the trap) to the arrangement on concentric rings characteristic of the surface region. 
\\ \indent
As mentioned above, the crystal to superfluid transition in the bulk is expected to take place roughly in the $0.045 \le r_s\le 0.065$ interval. The results obtained in this work show that a harmonically trapped system comprising at least $N$=1000 particles ought allow one to observe experimentally some of the relevant physics. On the other hand, the behavior of systems comprising just few hundred particles is still largely influenced by the confining potential, in the same range of $r_s$.  A more quantitative discussion of the physics of the trapped system, and it connection to the bulk superfluid to insulator quantum phase transition, is offered in {section~\ref{sec:phasediagram2}, based on the study of the superfluid properties of the system.

\section{Superfluidity and condensation}\label{sec:sfbec}

{Figure}~\ref{fig:rhosrhoc} shows a typical result for the superfluid and condensate fractions of a dipolar assembly. In this particular case, the number of particles $N$ is 100, and the trap strength {yields a value of $r_s$ in the middle of the trap  approximately equal to 0.22, }
as a function of temperature. As the temperature is lowered below $T^* \lesssim 1$, both the superfluid and condensate fractions become non-zero, saturating to very different values in the $T^\star \to 0$ limit. Specifically, the superfluid fraction $\rho_s$ approaches unity, whereas the condensate fraction $\rho_c$ is less than 20\% at $T^\star = 0$. On the other hand, in the large $r_s$ limit (weaker trap strength) superfluid and condensate fractions are {close} in magnitude, both close to unity at low temperature, as expected for a more \emph{dilute} system. 
That inter-particle interactions have a stronger (depleting) effect on the condensate than the superfluid fraction, is of course not a novel feature of this many-body system. The same trend is predicted theoretically \cite{moroni} and observed experimentally  \cite{glyde}  in condensed helium.

 \begin{figure}[t!]
\centering
\includegraphics[scale=1.1]{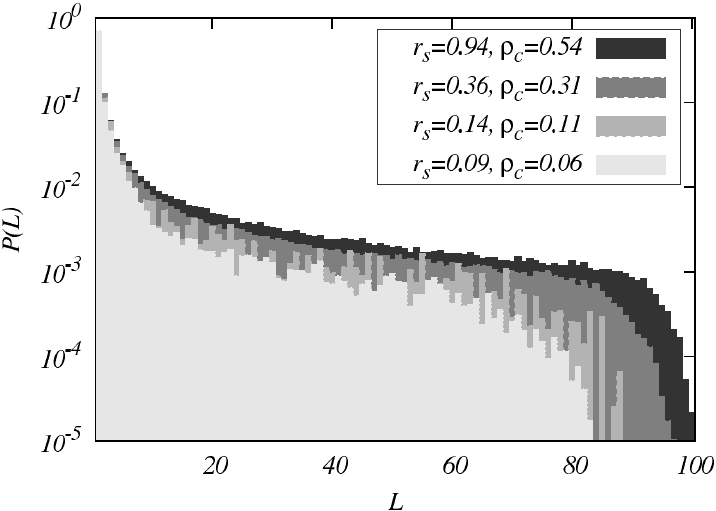}
\caption{Relative frequency of permutation cycles of length $L$ for simulations with four different trap strengths {corresponding to $r_s = 0.94, 0.36, 0.14, 0.09$} and for $N = 100$ and temperature $T^*=0.1$, at which the system is essentially in the ground state.}
\label{fig:pcycles}
\end{figure}

\begin{figure}[!t]
\centering
\includegraphics[scale=0.35]{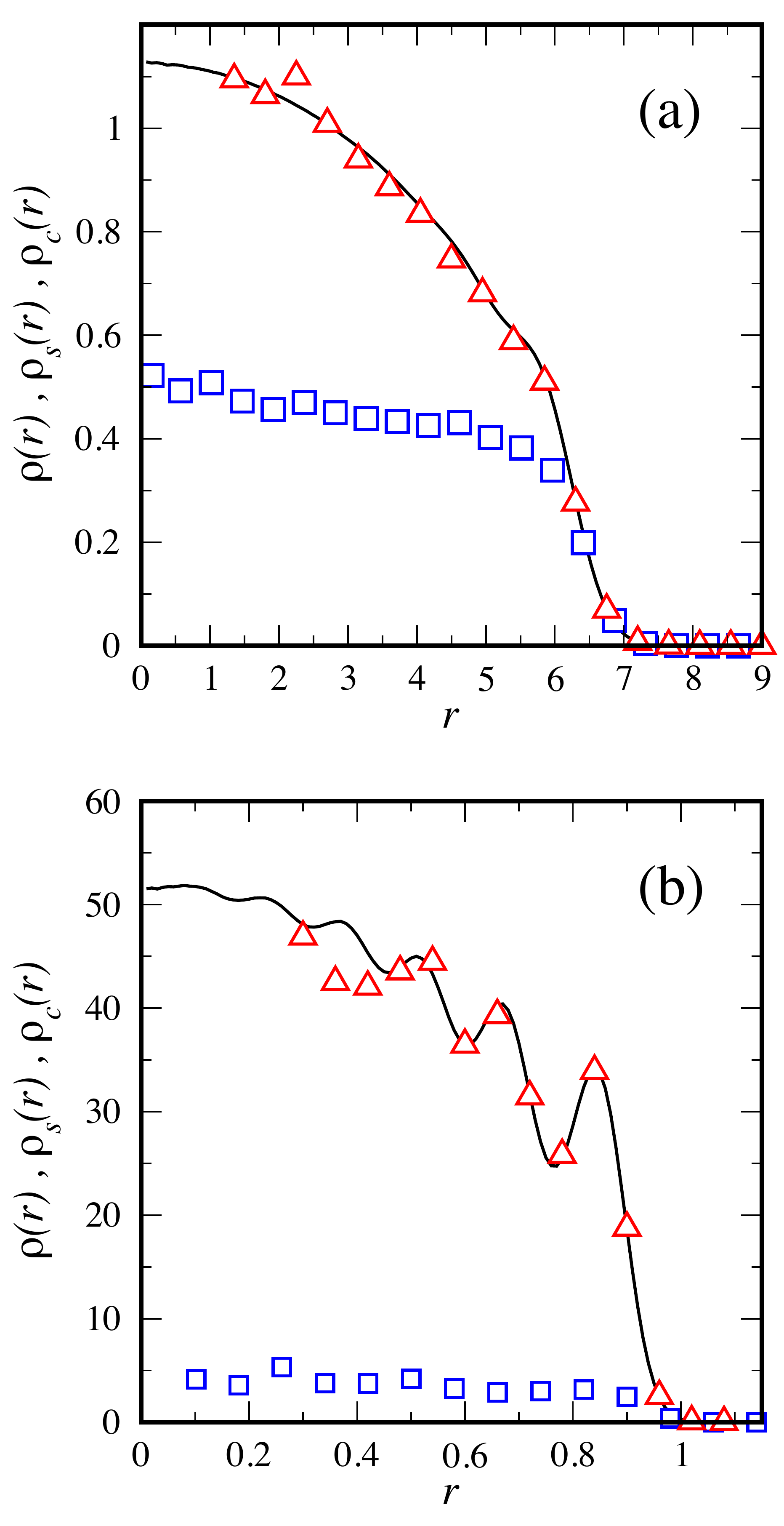}
\caption{(Color online) Total ($\rho(r)$, line),   condensate ($\rho_c(r)$, squares) and superfluid  ($\rho_s(r)$, triangles)  radial density profiles, for a system with $N$=100 dipolar bosons for two different trap strengths, yielding values of the inter-particle distance  $r_s$ in the center equal to 0.94, and  0.14. The system is in the ground state  and the overall superfluid fraction is near unity in both cases. Statistical and systematic uncertainties are estimated to be of the order of the sizes of the symbols.}
\label{fig:lsdlcd}
\end{figure}

It is well understood theoretically, that SF and BEC are both underlain by quantum-mechanical exchanges of identical particles \cite{ceperley}.  In order to explore this notion in greater detail, in Figure~\ref{fig:pcycles} we show the relative frequency of occurrence of exchange cycles involving a variable number $1 \le n \le N$ of particles (here, $N =100$). These results are for four different values of the  inter-particle distance in the center of the trap, namely {$r_s = 0.94, 0.36, 0.14, 0.09$} at low temperature. In all four cases, the superfluid fraction is 100\%, 
(within statistical uncertainties), whereas the condensate fraction goes from 54\% {for the the most dilute, to just 6\% (i.e., a mere {\it six} particles) for the densest assembly}. This difference in the fraction of the system condensed to the same single-particle quantum-mechanical state  is directly related to the much greater frequency with which  permutation cycles (of {\it any} length) occur at lower density, i.e., with weaker interactions. At the same time, these results clearly show that there is no obvious, direct {\it numerical} connection between the length of the longest observed exchange cycle, or the statistics of occurrence of cycles comprising specific numbers of particles, and the {superfluid} fraction. Remarkably, the superfluid signal in the low temperature limit is unaffected by the significant differences in the statistics of exchange cycles at the four thermodynamic conditions considered. 

As explained in section~\ref{sec:model},  we can compute local superfluid and condensate density, gaining greater insight into the relationship between SF and BEC.  Figure~\ref{fig:lsdlcd} shows the total,   superfluid   and  
condensate radial density profiles, for a system of $N$=100 particles and two different values of inter-particle separation $r_s$ in the center of the trap. The system is essentially in the ground state, i.e.,  the superfluid fraction  $\rho_s \sim 1$ {in both cases}. As one can see, the  condensate density  becomes suppressed relative to the  superfluid one, as 
the {inter-particle separation is decreased}. Significantly, however, both local quantities are almost uniformly distributed throughout the system. For the superfluid density, this is consistent with recent findings for parahydrogen clusters \cite{Mezzacapo2008}. The condensate density displays a markedly different beahvior in the vicinity of the surface, depending on the degree of particle localization. At the largest average inter-particle separation (lowest density), the system near the surface is highly dilute, and the local condensate fraction approaches 100\%,  as observed in helium droplets as well \cite{lewart}. As the density of the system is increased, particles near the surface become increasingly localized, and BEC is locally suppressed, while remaining finite near the center of the trap, where the physical behavior is that of a superfluid. 

It is worth noting that the behaviour of the system of dipolar bosons studied here deviates significantly from that of hardcore bosons studied previously by DuBois and Glyde \cite{DuBois2001}. In that study, it was found that the local condensate density  becomes severely depleted at the center of the trap, as the density is increased. In particular, the condensate  forms a shell at the edge of the trap for strongly interacting systems. This effect is absent here, a fact that can presumably be attributed to the long-range nature of the dipole-dipole interaction, {which in turn renders the use of any local density approximation questionable}. 

{As mentioned above and as will be quantitatively discussed in the next section, on decreasing  the average inter-particle separation a crystalline phase eventually nucleates in the center of the trap, at which point both superfluidity and condensation become suppressed in the system. The absence of a sharp phase transition in a finite system results in a value of the superfluid fraction less than one in the ground state, as superfluid and crystalline phases coexist. In the next section, we determine the ground state phase diagram of the system by calculating the superfluid and condensate fractions as a function of system size and inter-particle separation.}

\section{Ground state phase diagram}\label{sec:phasediagram2}

\begin{figure}[!tp]
\centering
\includegraphics[scale=0.35]{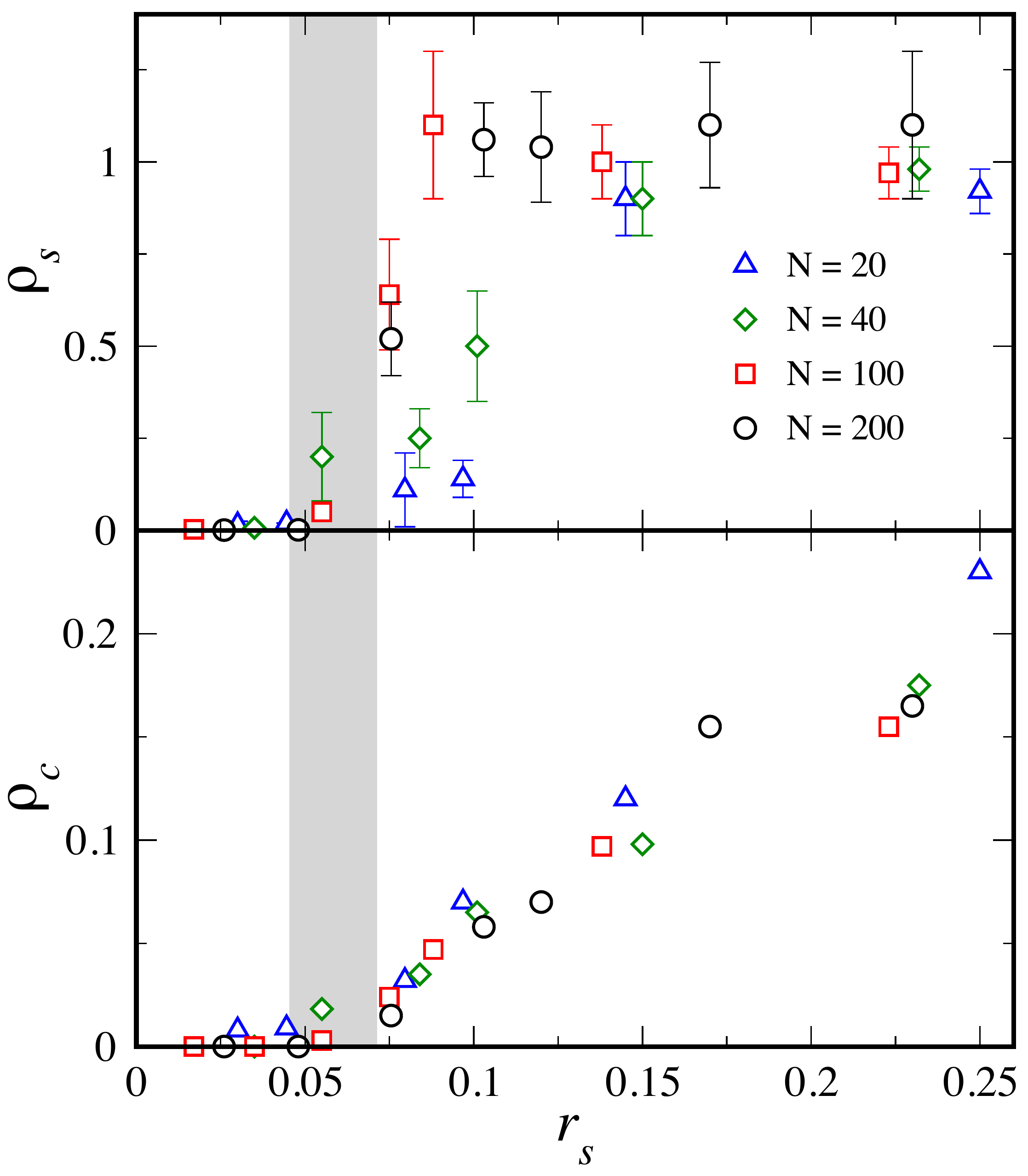}
\caption{(Color online) Top: Superfluid fraction (extrapolated to $T\rightarrow0$) $vs.$ interparticle distance $r_s$ for different system sizes. 
Bottom:
Condensate fraction $\rho_c$ as defined in the text (extrapolated to $T\rightarrow0$) $vs.$ interparticle distance $r_s$ for different system sizes. We have offset the condensate value by $1/N$, namely the smallest possible eigenvalue of the one-body density matrix. The shaded region represents the estimated range for the critical value of $r_s$ predicted for the homogeneous system from Ref. \cite{Buchler2007}}
\label{rhosrhoc}
\end{figure}

{As discussed in Sec.~\ref{sect:intro}}, a $T$=0 (quantum) phase transition between a crystalline, insulating phase  and a superfluid {phase} is observed in the bulk system, driven by the chemical potential (i.e., the density). One of the issues of interest here is, whether any remnant of such a transition can be observed, either by simulation or possible experiments, in a finite system. 
In Figure \ref{rhosrhoc} (top), we show values of the superfluid fraction $\rho_s$ corresponding to different interparticle distances $r_s$ in the middle of the trap. These values are computed for various system sizes (between $N=20$ and $N=200$), at a temperature sufficiently low to be regarded as ground state estimates.  
\\ \indent
Although no sharp phase transition can be observed in a finite system, the data   clearly show a crossover between two regimes, a superfluid  ($\rho_s \sim 1$) and crystal  ($\rho_s \sim 0$), taking place roughly in the same region of $r_s$ in which it is observed in numerical studies of the ground state bulk phase diagram \cite{Buchler2007} (shaded region of Fig. \ref{rhosrhoc}). It is worth mentioning that,  in the range $0.067\lesssim r_s\lesssim 0.12$, during a sufficiently long simulation the system switches back and forth between two different regimes, characterized by zero and finite superfluid density. This behavior, already reported in previous works \cite{Boning2008}, is not unlike what observed in small clusters of {para}hydrogen \cite{Mezzacapo2006,Mezzacapo2007}; it suggests that the two phases are energetically very close, and in practice renders lengthy simulations necessary, in order to assign a reliable value to the superfluid fraction. This is the reason for the (relatively) large statistical uncertainties of some of the estimates reported in Figure \ref{rhosrhoc}.
\\ 
\indent
The results in Figure \ref{rhosrhoc} (top) suggests that melting and freezing values for $r_s$ at $T$=0 should lie in the range 0.055 $\le r_s\le$ 0.07, which is consistent with bulk studies \newtext{\footnote{\newtext{In particular, in two separate quantum Monte Carlo studies of the bulk system the transition was predicted to occur either in the region $0.045 \leq r_s \leq 0.071$ \cite{Buchler2007} or in the region $0.056 \leq r_s \leq 0.062$ \cite{Astrakharchik2007}, in the latter case by extrapolating the system energy to the thermodynamic limit.}}}. A precise determination of the coexistence region is obviously beyond the scope of what can be achieved in a study of a finite system.




{Figure \ref{rhosrhoc} (bottom) shows the condensate fraction $\rho_c$ for the same set of simulation parameters for which the superfluid fraction is given in the top part. Here, we define ``condensate fraction" $\rho_c=N_0/N$, i.e., the largest eigenvalue $N_0$ of the cylindrically averaged one-particle density matrix, divided by the total number $N$ of particles. In order to facilitate the comparison of results obtained for systems of different size, we have subtracted $1/N$ from all estimates of $\rho_c$, as 1 is the lowest possible value of $N_0$. Such a correction is non-trivial for the smallest systems. 
\\ \indent
In the thermodynamic limit, $\rho_c = 0$ at any finite temperature, as thermal fluctuations destroy off-diagonal long-range order. In a finite system, however, $\rho_c$ can be finite, and indeed it is found to be finite for values of $r_s$ above $\sim$ 0.07, i.e., in the superfluid region. Remarkably, the data in Figure \ref{rhosrhoc} (bottom) are consistent with  the vanishing of $\rho_c$ concurrently with that of the superfluid fraction $\rho_s$, as the system is compressed to the crystalline phase. 



\begin{figure}[!t]
\centering
\includegraphics[scale=0.3]{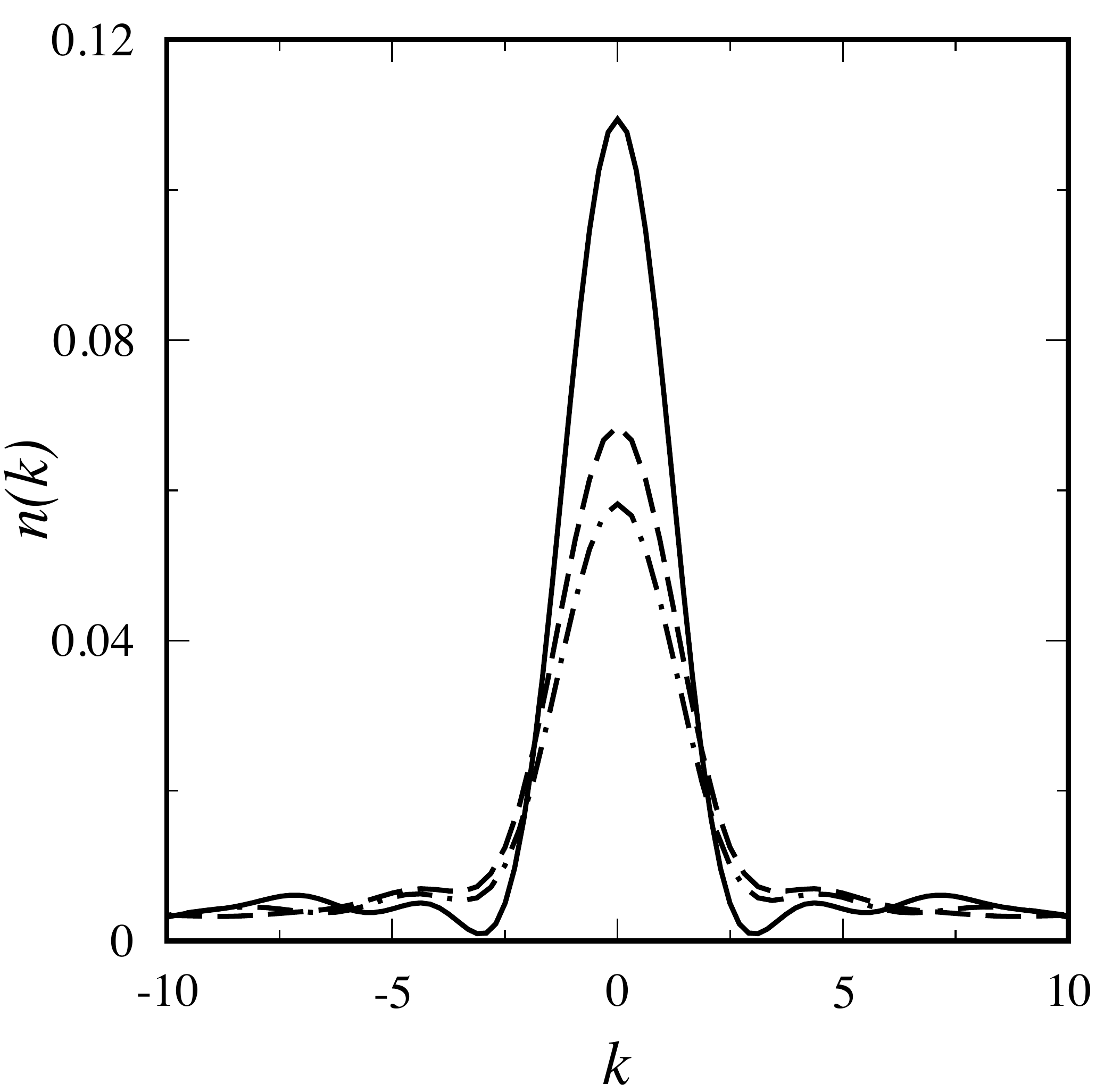}
\caption{(Colors online) Momentum distribution $n(k)$ for an assembly of $N=200$ dipolar bosons in a two-dimensional harmonic trap. The strength of the trap is such that the interparticle distance at its center is $r_s=0.122$. 
Results shown are for temperatures $T$=0.05 (solid line), $T=0.2$ (dashed line), and $T=0.4$ (dot-dashed line), all in units of $T^\star$, defined in the text. }
\label{figure4}
\end{figure}




In an infinite system, the single-particle state in which bosons condense is a plane wave of momentum ${\bf k}=0$, simply based on considerations of translational invariance. This results in a (theoretical) $\delta$-like peak at ${\bf k}=0$ in $n({\bf k})$. Experimental evidence of phase coherence throughout the system can be detected in the $n({\bf k})$ of a spatially confined system as well, through the appearance of a peak at ${\bf k}=0$ at low temperature, even though harmonic confinement has the effect of broadening it, its width being of order 1/$L$. \\ 
In experiment with trapped cold atoms, the momentum distribution is  routinely measured via time-of-flight absorption imaging.  As a guide for possible experiments, it is therefore instructive to compute the momentum distribution  in the limit $T^* \rightarrow 0$,  for two systems which have a superfluid and a crystalline ground state.
\\
\indent
For a two-dimensional cylindrically symmetric system, the momentum distribution is given by 
\begin{equation}
n(k) = \int^{\infty}_0 \ dr\ r\ {\tilde n}(r)\ J_0(k r) 
\end{equation}
 where ${\tilde n}(r)$ is the spatially averaged one-body density matrix, computed by simulation,  and $J_0$ is the zeroth order Bessel function \cite{Boninsegni2009}.
Figure \ref{figure4} shows the computed $n(k)$ for a system of $N$=200 dipolar bosons at three different temperatures, for a trap strength yielding a value of $r_s\sim 0.122$ in the middle of the trap, i.e., well into the superfluid region of the phase diagram. The development of a sharp peak at $k$=0 as the temperature is lowered is clear; the value of the superfluid fraction at low $T$ approaches 100\% in this case.  No evidence of such a peak is observed in our simulations for values of $r_s$ for which the assembly of dipolar particles does not display a measurable superfluid response.
\\
Thus, the results of this study suggest that much of the physics of the crystal-superfluid transition in dipolar bosons can be probed experimentally on a system comprising as few as several hundred particles.

\section{Discussion and Conclusion}\label{sec:conclusions}

We have investigated by extensive {Quantum Monte Carlo} simulations the properties of
dipolar bosons confined in  two-dimensional traps.
In particular we have analyzed {the structure of the system, superfluidity and Bose-Einstein condensation, as well as the ground state phase diagram for the} Hamiltonian~\eqref{hamiltonian} for different system sizes and as a function of
in-trap density. {We have considered system sizes up to $N=1000$, a number relevant to current and planned experiments. It is therefore our expectation that a direct comparison between theory and experiment should be possible.} 

We find that the physics of the system in the middle of the trap closely mimics that expected in an infinite system, at least in the region of density wherein the superfluid-insulator quantum phase transition is expected, if the number of particles in the trap is of order of a few hundreds.  In particular, a transition between superfluid and crystal like phases is evident, as shown by our results for the density profiles, local pair correlation function as well as the superfluid and condensate fractions. {By considering system sizes up to $N=1000$, we verify that such a transition occurs for an inter-particle separation of 0.055 $\le r_s\le$ 0.07, which is consistent with the most reliable theoretical predictions for the bulk case \cite{Buchler2007}.
}


In order to gain deeper understanding of the elusive relationship between superfluidity 
and Bose-Einstein condensation, a study of the global and local superfluid properties  
has also been performed. The general conclusion is consistent with that obtained for systems with different interactions, namely that strong inter-particle interactions have a much more pronounced quantitative effect on the condensate than on the superfluid fraction. However, our results are consistent with the concurrent disappearance of both as the system freezes at high density.  With the exception of a narrow region near the edge of the trap, where BEC is either enhanced or suppressed depending on particle localization, superfluid and condensate density are largely uniform throughout the system.  

In striking contrast from the case of hardcore bosons \cite{DuBois2001}, 
BEC is not depressed at the center of the trap in the presence of strong interactions (i.e., at high density).
Such a qualitative difference points to the importance of the long range nature of the interactions in \eqref{hamiltonian}, with the inference that a local density approximation cannot be straightforwardly applied here.

We have also investigated the momentum distribution  as a function of temperature. The presence of coherence at low temperature, if the system features a superfluid ground state, is rendered evident by the appearance of a pronounced peak at zero momentum, in the low temperature limit. This seems to indicate that important signatures of a quantum phase transition predicted in the bulk system, should be readily observed in current or planned experiments.\\


\section*{Acknowledgements}
The authors thank Fabio Mezzacapo for useful discussions.
This work was supported in part by the Natural Science and Engineering 
Research Council of Canada under research grant 121210893, and by the 
Alberta Informatics Circle of Research Excellence (iCore).

\bibliography{references}

\end{document}